# Predicting Toxicity from Gene Expression with Neural Networks

Peter Eastman[1] and Vijay S. Pande[1]

[1]Department of Bioengineering, Stanford University, Stanford, CA 94305

## Abstract

We train a neural network to predict chemical toxicity based on gene expression data. The input to the network is a full expression profile collected either *in vitro* from cultured cells or *in vivo* from live animals. The output is a set of fine grained predictions for the presence of a variety of pathological effects in treated animals. When trained on the Open TG-GATEs database it produces good results, outperforming classical models trained on the same data. This is a promising approach for efficiently screening chemicals for toxic effects, and for more accurately evaluating drug candidates based on preclinical data.

## Introduction

Predicting toxicity is a vital problem in many fields. One quarter of all drug candidates that reach phase II clinical trials ultimately fail because of toxicity[1]. Better methods to predict this in advance would spare patients from taking drugs that ultimately prove toxic, as well as saving enormous time and money. Toxic effects from industrial and household chemicals are also a major public health problem. Often they are tested only on animals, not humans, but animals can be a poor model for toxicity in humans[2]. Better methods to predict human toxicity would have major public health benefits.

Many chemicals cause chronic rather than acute toxicity. It may take months or years for their effects to become apparent. Better methods to spot the early signs of chronic toxicity before clinical symptoms appear would allow clinical trials to be stopped sooner, and also would reduce the risk of toxic effects being missed.

Gene expression data is often used for predicting toxicity, since it is easy to obtain and highly informative[3]. Usually the prediction is made by an indirect, multistep process. 1. Identify the most highly differentially expressed genes between treated and control samples. 2. Compare the list of genes to standard ontologies to identify biological pathways affected by the chemical[4]. Alternatively, one may compare to lists of genes known to be affected by other compounds[5,6], then assume the new compound probably has similar effects to ones that affect similar genes. Either way, the results are imprecise and require substantial human interpretation.

The process also loses information. The original expression data includes one number per gene per sample, but the list of differentially expressed genes includes only a single bit per gene, a huge reduction in information content. This suggests better results could possibly be obtained by methods that directly analyze the full expression profile.

Several recent studies attempted to do this, with significant success[7–9]. These studies relied on relatively simple machine learning models such as Support Vector Machines (SVMs) and naive Bayes classifiers. They also classified chemicals into only a few coarse grained categories (e.g. "toxic" and "nontoxic"), rather than attempting fine grained predictions of the many types of toxicity that can occur.

In recent years, deep learning has achieved great success in fields such as computer vision and natural language processing, largely replacing classical machine learning models in those fields[10]. This leads us to speculate that deep neural networks might also be effective for analyzing gene expression data. Our goal in this work is to create a neural network that takes gene expression profiles as input and outputs predictions of toxic effects.

We train and test it with the Open TG-GATEs database[11]. This database contains data on 170 compounds that were tested *in vivo* on rats, as well as *in vitro* on rat and human liver cells. A full expression profile was collected for every experiment. For the *in vivo* experiments, liver and kidney samples were visually examined and graded on each of several dozen histopathological findings. Our model attempts to directly predict these pathological findings based on either *in vivo* or *in vitro* expression data. As described later, our results are highly encouraging.

This study is intended as a proof of concept. Ultimately, of course, we would like to predict *in vivo* toxic effects in humans, but given the available public datasets, we settle for predicting them in rats. With appropriate data to train on, the method should work equally well for any species.

# Results

## Prediction from *in vivo* data

We first build a neural network that takes the expression profile of an *in vivo* sample as input, and outputs a number for each of the 15 most common pathological findings. Open TG-GATEs rates each finding as one of six grades. In order of increasing severity, they are: not present, present, minimal, slight, moderate, and severe. We map these to the integers from 0 to 5 and use them as labels for the model to predict. This means it is a regression model, not a classification model. It tries to predict the severity of each finding, not just whether it is present. This also eliminates the need for ambiguous judgements about which compounds to label as "toxic". For each sample, the model tries to predict which pathologies were actually present in that particular sample.

The distribution of test set samples is shown in Figure 1 for each pathological finding. Nearly all of them show a clear monotonic relationship between measured and predicted severity. The model tends to output higher values for samples that actually displayed the pathology, and the output value increases with increasing severity. The sole exception is microgranuloma, which shows no correlation between true and predicted severities.

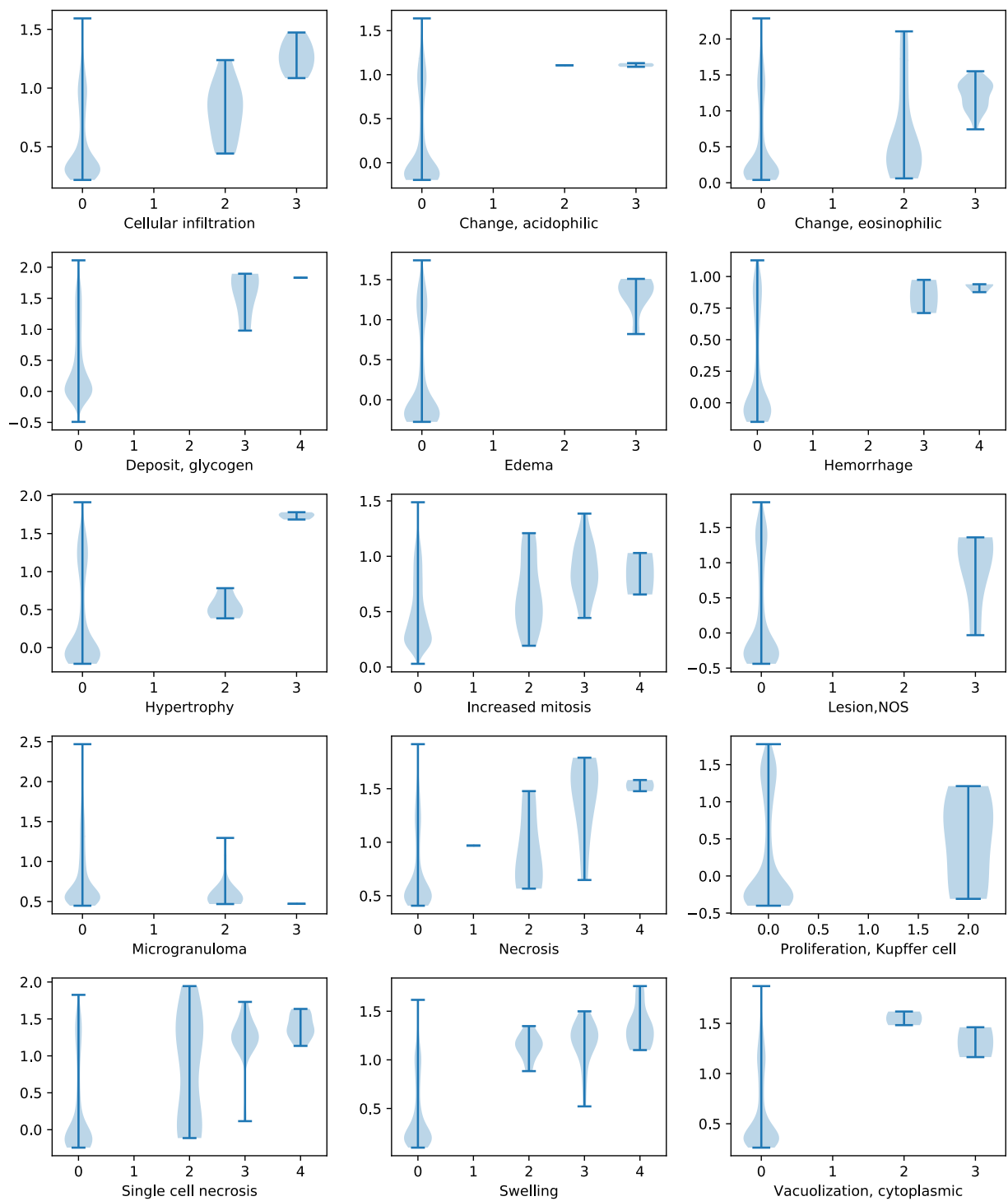

Fig 1. Distribution of predicted severity (vertical axis) versus experimental severity (horizontal axis) for all samples in the test set for each of the predicted pathological findings.

In addition to the neural network, we also trained two classical models on the same data: a linear SVM and a random forest (RF). Table 1 shows the weighted $L_2$ error (which was used as the loss function during training) for all three models. The neural network easily outperforms the classical models, having the lowest error on 13 of the 15 prediction tasks. This shows its predictions are significantly more accurate.

|  | Loss | | | ROC-AUC | | |
| --- | --- | --- | --- | --- | --- | --- |
|  | NN | SVM | RF | NN | SVM | RF |
| Cellular infiltration | 1088 | **1033** | 1999 | 0.973 | **0.988** | 0.941 |
| Change, acidophilic | **1438** | 3072 | 3766 | 0.930 | **0.971** | 0.686 |
| Change, eosinophilic | **1660** | 2735 | 2895 | 0.818 | 0.712 | **0.855** |
| Deposit, glycogen | **1447** | 5172 | 4859 | **0.939** | 0.647 | 0.860 |
| Edema | **1366** | 1990 | 2820 | 0.917 | **0.983** | 0.906 |
| Hemorrhage | **3707** | 5567 | 6318 | 0.883 | **0.924** | 0.720 |
| Hypertrophy | **1045** | 1752 | 2132 | **0.995** | 0.961 | 0.980 |
| Increased mitosis | **1981** | 2826 | 3151 | 0.815 | 0.804 | **0.906** |
| Lesion, NOS | **2684** | 4548 | 5071 | 0.761 | **0.812** | 0.760 |
| Microgranuloma | **1301** | 2230 | 2092 | - | - | - |
| Necrosis | **1119** | 1549 | 2107 | **0.915** | 0.886 | 0.864 |
| Proliferation, Kupffer cell | **1733** | 2313 | 2435 | - | - | - |
| Single cell necrosis | 1988 | **1611** | 2423 | 0.865 | **0.971** | 0.965 |
| Swelling | **2806** | 6873 | 6793 | **0.952** | 0.421 | 0.539 |
| Vacuolization, cytoplasmic | **698** | 1735 | 1630 | 0.916 | 0.912 | **0.925** |

Table 1. Performance on the test set for each of the three models (neural network, support vector machine, random forest) on each of the predicted pathological findings. The best scoring model in each case is shown in bold. ROC-AUC was not computed for two of the findings because there were not sufficient test samples rated as "slight" or higher.

Although the models were trained as regression models, it is also informative to see how well they can perform classification. For each pathology we divide the samples into two groups: those rated "minimal" or lower, and those rated "slight" or higher. This division is motivated by observing that even untreated samples often display pathologies at the "minimal" level, so "slight" is the lowest level reliably indicating a toxic effect.

Table 1 shows the ROC-AUC scores (area under the receiver operating characteristic curve). The neural network again performs best, with a mean AUC over all pathologies of 0.898. It is followed by the SVM with a mean AUC of 0.846, and the RF with a mean AUC of 0.839.

## Prediction from *in vitro* data

We next try to predict pathological findings based on *in vitro* expression data. Experiments on cultured human cells are inexpensive to perform, and they avoid the many ethical concerns of testing potentially toxic chemicals on humans. An effective prediction model that requires only preclinical data would allow better decisions about which drug candidates should advance to clinical trials. It also would enable fast, inexpensive screening of industrial chemicals for potential toxicity.

Since the input values (expression levels) and labels (pathological findings) now come from different experiments, some care is required in matching them up. The *in vivo* and *in vitro* data sets both include multiple dose levels, and the correspondence between them is ambiguous. If a pathological finding was observed at "slight" or higher level in at least two treated *in vivo* samples (any dose) for a compound, the compound is considered to potentially cause that pathology. We use only the highest dose level *in vitro* samples, since those are most likely to actually display a toxic effect. Our neural network takes an *in* *vitro* expression profile as input, and predicts whether or not the sample was treated with a drug that causes each pathology *in vivo*.

Clearly this process is imperfect. Even if a chemical sometimes causes a particular toxic effect, that does not guarantee the effect will actually be present in every treated sample. Our hope is that even when no toxic effect is observed, gene expression will still be disrupted enough for the model to recognize that a drug *could* cause a particular effect. Likewise, when a toxic effect takes time to develop, we hope that warning signs of it will be detectable in expression profiles from earlier time points.

As before we train three models: a neural network, a linear SVM, and a random forest. The ROC-AUC scores are shown in Table 2. The neural network easily outperforms both of the classical models with a mean AUC over all pathologies of 0.754, compared to 0.627 for the random forest and only 0.543 for the SVM. In fact, the SVM has minimal predictive ability (AUC below 0.6) for all but one task. The random forest does somewhat better, but it still has minimal predictive ability on four of the nine tasks. In contrast, the neural network has AUC scores above 0.6 for all but one task. The one task on which all models fail, "Deposit, pigment", happens to be the task with the least training data, having only 12 positive examples in the training set. There may simply not be sufficient data for any of the models to learn this task.

|  | ROC-AUC | | |
| --- | --- | --- | --- |
|  | NN | SVM | RF |
| Change, eosinophilic | **0.802** | 0.500 | 0.481 |
| Deposit, pigment | 0.451 | 0.500 | 0.522 |
| Ground glass appearance | **0.690** | 0.583 | 0.436 |
| Hypertrophy | **0.858** | 0.583 | 0.762 |
| Increased mitosis | **0.667** | 0.663 | 0.520 |
| Proliferation, bile duct | **0.972** | 0.500 | 0.747 |
| Single cell necrosis | 0.759 | 0.481 | **0.791** |
| Swelling | **0.937** | 0.500 | 0.683 |
| Vacuolization, cytoplasmic | 0.651 | 0.583 | **0.705** |

Table 2. Performance on the test set for each of the three models (neural network, support vector machine, random forest) on each of the predicted pathological findings. The best scoring model in each case is shown in bold, except in the one case where all models had negligible predictive ability.

The inferior performance of the SVM and RF models seems to be caused primarily by overfitting. All three models do quite well at predicting the training set (data not shown), with most AUC scores being above 0.98. But they do not generalize well to the test set. Possibly their performance could be improved through regularization, although random forests and SVMs are both designed to be inherently resistant to overfitting. The neural network does a better job of generalizing to data that was not used during training.

Figure 2 shows the neural network's ROC curve (true positive rate versus false positive rate) for each of the nine pathologies. On some of the tasks ("change, eosinophilic", "proliferation, bile duct", and "swelling"), the true positive rate saturates to 1.0 when the false positive rate is still at a quite low value. This makes the model particularly effective in cases when it is important to correctly classify all toxic compounds, and a moderate number of false positives is acceptable. For example, this would be the case when using the model to screen industrial chemicals and identify ones requiring further study.

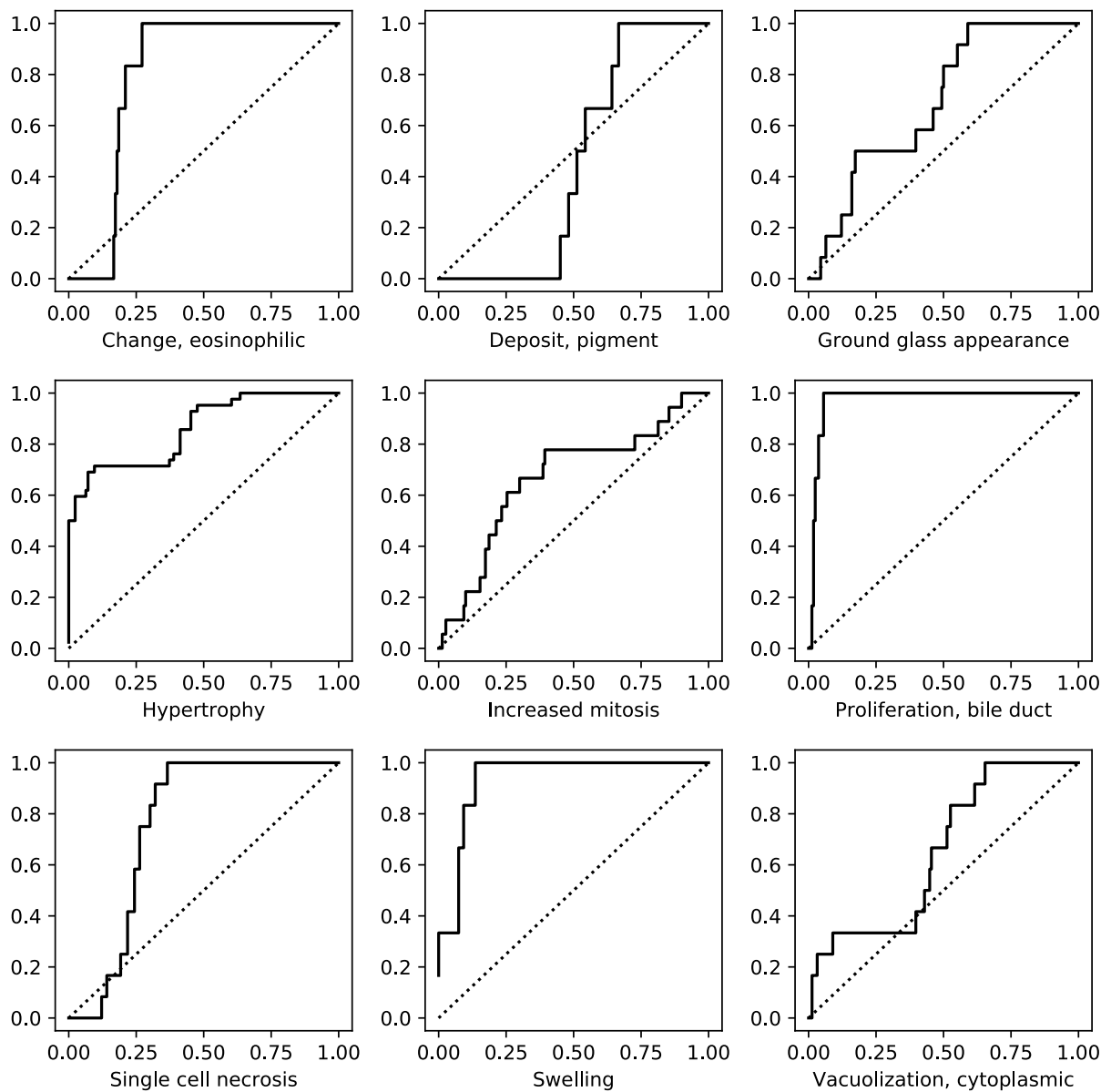

Fig 2. ROC curve (true positive rate versus false positive rate) for the neural network on each of the predicted pathological findings. Wherever the line is above the diagonal, that indicates the model is doing better than would be expected by chance.

# Discussion

In this work we train a neural network to directly predict toxicity from a complete gene expression profile. Compared to previous studies and to the standard techniques in the field, our method has several important benefits.

Our model produces explicit predictions about whether or not compounds are toxic. This is qualitatively different from the conventional methods, whose outputs are lists of affected biological pathways, or alternatively lists of known compounds with similar effects on gene expression. This makes it complementary to them. For understanding a mechanism of action, a list of affected pathways is very valuable information, but it also is complex information that requires much human interpretation. In many situations it is valuable to have explicit predictions about whether a compound is or is not toxic.

Our model operates directly on expression profiles rather than on lists of differentially expressed genes. This gives it the potential to be more accurate. The transition from expression levels to a list of genes involves tremendous loss of information. In some cases, small changes in the expression of many genes may be a better predictor than large changes in just a few genes, but conventional methods cannot detect it.

Compared to other studies that have tried to directly predict toxicity from expression, our model's predictions are much more detailed and fine grained. For example, in Liu et al.[7] each compound was simply labeled as toxic or non-toxic for each of the studied organs. In Takahashi et al.[9] each compound was placed into one of three categories: genotoxic carcinogens, non-genotoxic carcinogens, and neurotoxins. In contrast, our model independently predicts many different pathological features, thus giving more detailed information about the likely effects of a compound.

We use a neural network, in contrast to the simpler classical models used in other studies. In our tests this leads to a substantial improvement in the accuracy of predictions.

We see two primary applications for this method. First, it can be used to predict clinical toxicity based on preclinical data collected from cultured cells. This will allow a fast, inexpensive assay to screen for toxic effects and identify chemicals requiring further study, or promising drug candidates for human trials. Second, it can be used to identify early signs of chronic toxicity in either clinical or preclinical studies. This will allow faster and more reliable identification of toxicity, thus saving time, money, and human suffering. Our results are very encouraging for its ability to perform these functions.

The main challenge in applying this method will be the creation or identification of appropriate data sets for training. In this work we apply it to predicting toxicity in rats because of the public availability of high quality data. Ultimately, of course, our goal is to predict toxicity in humans. The model is completely generic with respect to species and should work just as well for humans as for rats. To do that, it needs a large collection of expression data from normal and pathological human samples, along with high quality labels for any pathologies present in each sample.

# Methods

## Model for *in vivo* data

The data set consists of all single-dose *in vivo* rat liver samples from the Open TG-GATEs database[11]. This is a total of 7378 samples treated with 158 compounds. Each sample is graded on several dozen pathological findings, but most findings are only observed for a very small number of samples, or in many cases never observed at all. To ensure sufficient data for training and testing, we use only the 15 most common findings, shown in Table 3. Each of them is present in at least ten samples, and is observed for at least three different compounds.

| Pathological Finding | Training Samples | Test Samples | Training Compounds | Test Compounds |
|---|---|---|---|---|
| Cellular infiltration | 112 | 19 | LPS, acetaminophen, allyl alcohol, bendazac, bromobenzene, carbamazepine, carbon tetrachloride, cimetidine, clofibrate, ethionine, galactosamine, glibenclamide, griseofulvin, hexachlorobenzene, isoniazid, lornoxicam, methylene dianiline, phorone, sulfasalazine, terbinafine, valproic acid | acetamidofluorene, naphthyl isothiocyanate, thioacetamide |
| Change, acidophilic | 10 | 3 | N-nitrosomorpholine, cyclosporine A | nitrosodiethylamine |
| Change, eosinophilic | 90 | 47 | LPS, acetaminophen, aflatoxin B1, bortezomib, chlormezanone, coumarin, diethyl maleate, ethionamide, ethionine, nicotinic acid, omeprazole, phorone | cycloheximide, phenylanthranilic acid, thioacetamide |
| Deposit, glycogen | 20 | 5 | chlormadinone, nicotinic acid | bromoethylamine |
| Edema | 45 | 7 | LPS, methylene dianiline | naphthyl isothiocyanate |
| Hemorrhage | 22 | 5 | methylene dianiline, sulfasalazine | phalloidin |
| Hypertrophy | 39 | 8 | LPS, bortezomib, bromobenzene, ethionamide, imipramine, methapyrilene, methylene dianiline, methyltestosterone, puromycin aminonucleoside, sulindac | benzbromarone, thioacetamide |
| Increased mitosis | 92 | 39 | 3-methylcholanthrene, WY-14643, bendazac, benziodarone, coumarin, diethyl maleate, ethambutol, ethinylestradiol, ethionamide, fenofibrate, flutamide, gemfibrozil, griseofulvin, ibuprofen, ketoconazole, methyltestosterone, nifedipine, papaverine, phorone, propylthiouracil, simvastatin | colchicine, danazol, nimesulide, phenylanthranilic acid |
| Lesion,NOS | 9 | 5 | benziodarone, gefitinib | imatinib, methanesulfonate salt, phalloidin |
| Microgranuloma | 73 | 19 | acarbose, ajmaline, bendazac, benziodarone, buthionine sulfoximine, captopril, chloramphenicol, cimetidine, ciprofloxacin, desmopressin acetate, ethinylestradiol, ethionamide, etoposide, fluoxetine hydrochloride, flutamide, glibenclamide, lornoxicam, tetracycline | colchicine, ethanol, nimesulide |
| Necrosis | 174 | 20 | 3-methylcholanthrene, LPS, WY-14643, acarbose, ajmaline, allopurinol, allyl alcohol, amiodarone, amitriptyline, amphotericin B, aspirin, bendazac, benziodarone, bortezomib, bromobenzene, buthionine sulfoximine, captopril, carbon tetrachloride, chloramphenicol, chlormezanone, chlorpromazine, ciprofloxacin, clofibrate, dantrolene, desmopressin acetate, diclofenac, diethyl maleate, diltiazem, ethinylestradiol, ethionamide, ethionine, etoposide, fluoxetine hydrochloride, flutamide, furosemide, galactosamine, gemfibrozil, glibenclamide, griseofulvin, hexachlorobenzene, indomethacin, isoniazid, lomustine, metformin, methyldopa, methylene dianiline, nitrofurazone, omeprazole, pemoline, phenobarbital, phorone, simvastatin, sulfasalazine, tannic acid, terbinafine, valproic acid | colchicine, imatinib, methanesulfonate salt, naphthyl isothiocyanate, nimesulide, nitrosodiethylamine, phalloidin, phenylanthranilic acid, tamoxifen, thioacetamide |
| Proliferation, Kupffer cell | 6 | 5 | chlorpropamide, puromycin aminonucleoside, theophylline | rosiglitazone maleate |
| Single cell necrosis | 142 | 63 | 3-methylcholanthrene, LPS, N-nitrosomorpholine, WY-14643, aflatoxin B1, bortezomib, coumarin, diethyl maleate, ethionamide, galactosamine, | colchicine, cycloheximide, nitrosodiethylamine, phalloidin, thioacetamide |

| | | | gefitinib, methapyrilene, methylene dianiline, phorone, tunicamycin | |
| Swelling | 36 | 45 | LPS, aspirin, isoniazid | acetamidofluorene, nitrosodiethylamine |
| Vacuolization, cytoplasmic | 54 | 4 | bendazac, bortezomib, dexamethasone, diethyl maleate, enalapril, ethionine, lomustine, phenylbutazone, phorone, rifampicin, simvastatin, tetracycline | colchicine, cycloheximide |

Table 3. The pathological findings predicted by the *in vivo* model. For each one, the table lists the specific training and test compounds for which the finding was observed, as well as the total numbers of training and test samples in which it was observed.

We divide the samples by compound into training (142 compounds) and test (16 compounds) sets. This ensures that all treated samples in the test set are for compounds not present in the training data. The choice of which compounds to include in each set is highly constrained by the requirement that every pathology be present in at least two training compounds and at least one test compound, so as to have sufficient data for training and testing.

The model is a fully connected neural network with two hidden layers of width 4000 and 2000, respectively[10]. The input is the normalized log scale expression values for 14,075 rat genes, and the output is a list of 15 numbers giving the predicted severities for the 15 pathological findings. All layers except the output use rectified linear unit (ReLU) activation and 50% dropout[12] during training. The model is trained to reproduce the measured severities (not present=0, present=1, minimal=2, slight=3, moderate=4, severe=5) using an Adam optimizer[13] with batch size 500 for 2000 epochs. The loss function is $L_2$ error, weighted so that the total weight of positive and negative samples is equal for each pathology. The learning rate is initially set to $5 \cdot 10^{-5}$, then decayed by multiplying it by 0.9 every 2000 steps.

## Model for *in vitro* data

The data set consists of all *in vitro* rat samples from Open TG-GATEs for which the dose is listed as either "Control" or "High". The list of compounds is restricted to those which were measured in both *in vitro* and *in vivo* experiments. This yields a total of 840 control samples and 830 treated samples for 140 compounds. We use the same 16 compounds as before for the test set, leaving 124 compounds for the training set.

As described previously, the identification of which compounds produce which pathologies is based on the observed findings for *in vivo* liver samples. If a pathological finding is present at "slight" or higher level in at least two treated samples (any dose) for a compound, that compound is considered to cause the pathology. We restrict the dataset to those pathologies caused by at least two training compounds and at least one test compound. This yields nine pathologies, shown in Table 4.

| Pathological Finding | Training Samples | Test Samples | Training Compounds | Test Compounds |
|---|---|---|---|---|
| Change, eosinophilic | 96 | 6 | acetaminophen, adapin, amiodarone, bucetin, chlormezanone, clomipramine, cycloheximide, disopyramide, ethambutol, ethinylestradiol, ethionamide, ethionine, nicotinic acid, omeprazole, phenylanthranilic acid, phorone | aspirin |
| Deposit, pigment | 12 | 6 | ethambutol, iproniazid | methapyrilene |
| Ground glass appearance | 36 | 12 | benzbromarone, dantrolene, ethambutol, hexachlorobenzene, nimesulide, phenobarbital | papaverine, promethazine |
| Hypertrophy | 222 | 42 | WY-14643, amitriptyline, bendazac, benzbromarone, benziodarone, bromobenzene, bromoethylamine, caffeine, carbamazepine, chlormezanone, chlorpropamide, coumarin, dantrolene, diazepam, disulfiram, erythromycin ethylsuccinate, ethambutol, flutamide, hexachlorobenzene, ketoconazole, lomustine, methimazole, methyltestosterone, mexiletine, monocrotaline, naproxen, nimesulide, omeprazole, phenacetin, phenobarbital, phenytoin, quinidine, terbinafine, thioacetamide, ticlopidine, tolbutamide, vitamin A | chloramphenicol, diltiazem, hydroxyzine, imipramine, methapyrilene, papaverine, promethazine |
| Increased mitosis | 78 | 18 | WY-14643, colchicine, diethyl maleate, ethambutol, ethionamide, fenofibrate, gemfibrozil, ibuprofen, ketoconazole, methyltestosterone, phorone, puromycin aminonucleoside, simvastatin | danazol, methapyrilene, papaverine |
| Proliferation, bile duct | 30 | 6 | acetamidofluorene, lomustine, naphthyl isothiocyanate, phalloidin, thioacetamide | methapyrilene |
| Single cell necrosis | 48 | 12 | colchicine, cycloheximide, ethambutol, ethionamide, galactosamine, phalloidin, theophylline, thioacetamide | methapyrilene, tunicamycin |
| Swelling | 36 | 6 | acetamidofluorene, clofibrate, griseofulvin, isoniazid, nitrosodiethylamine, sulfasalazine | aspirin |
| Vacuolization, cytoplasmic | 54 | 12 | amiodarone, amitriptyline, bromoethylamine, ethinylestradiol, ethionamide, hexachlorobenzene, ketoconazole, phorone, vitamin A | hydroxyzine, imipramine |

Table 4. The pathological findings predicted by the *in vitro* model. For each one, the table lists the specific training and test compounds for which the finding was observed, as well as the total numbers of training and test samples treated with those compounds.

The model is a fully connected neural network with two hidden layers of width 4000 and 2000, respectively. The input is the normalized log scale expression values for 14,075 rat genes, and the output is a list of nine pairs of numbers giving the predicted probability that each of the nine pathological findings is or is not present in the sample. All layers except the output use ReLU activation and 50% dropout during training. The output layer uses softmax activation to produce values that can be interpreted as probabilities. The model is trained to reproduce a set of labels that equal 1 if a sample was treated with a compound that causes the corresponding pathology, 0 otherwise. Training uses an Adam optimizer with batch size 500 for 5000 epochs. The loss function is the cross entropy between outputs and labels, weighted so that the total weight of positive and negative samples is equal for each pathology. The learning rate is initially set to $2 \cdot 10^{-5}$, then decayed by multiplying it by 0.9 every 1000 steps.

## Implementation

RMA normalization[14] of the expression data is performed with the pyAffy library[15] using the Rat2302_Rn_ENTREZG_22.0.0 CDF file from the BrainArray website[16]. The models are implemented with DeepChem 2.1[17], TensorFlow 1.7[18], and scikit-learn 0.19.1[19]. Training is performed on a NVIDIA Titan X GPU.